\documentclass[conference]{IEEEtran}
\pdfoutput=1	
\usepackage{booktabs} 
\usepackage{cite}

\usepackage{listings}
\usepackage{graphicx}
\graphicspath{ {.} {local_images/} {local_plots/} }
\usepackage[utf8]{inputenc}
\usepackage[linesnumbered,noend]{algorithm2e}
\usepackage[noend]{algpseudocode}
\usepackage{enumitem}
\usepackage{multicol}	

\usepackage[normalem]{ulem}
\usepackage{xcolor}    
\definecolor{darkGreen}{RGB}{0,153,0}

\usepackage{makecell}
\usepackage{indentfirst}		

\usepackage{epsfig}  
\usepackage[cmex10]{amsmath}  
\usepackage{amssymb}  
\usepackage{url}  
\usepackage{lscape}  
\usepackage[justification=raggedright]{caption}	
\usepackage{subcaption}

\usepackage[english]{babel}
\usepackage{verbatim}             
\usepackage[margin=0.75in]{geometry} 
\usepackage {multirow}
\usepackage{capt-of}		  

\usepackage{amsfonts}

\SetKwInput{KwData}{Initialize}
\SetKwInput{KwInput}{Input}                
\SetKwInput{KwOutput}{Output}              

\newcommand{\Id}[1]{I}
\newcommand{\Zr}[1]{0}

\newcommand{\edgar}[1]{{\it \color{red}Edgar: #1}}

\newcommand{\Var}{\operatorname{Var}}
\newcommand{\Expe}{\operatorname{E}}
\begin{document}

\title{Accelerating Distributed-Memory Autotuning via Statistical Analysis of Execution Paths}

\author{\IEEEauthorblockN{Edward Hutter}
\IEEEauthorblockA{\textit{Department of Computer Science}\\
\textit{University of Illinois at Urbana-Champaign}\\
hutter2@illinois.edu}
\and
\IEEEauthorblockN{Edgar Solomonik}
\IEEEauthorblockA{\textit{Department of Computer Science}\\
\textit{University of Illinois at Urbana-Champaign}\\
solomon2@illinois.edu}}

\maketitle

\begin{abstract}
\par{}
The prohibitive expense of automatic performance tuning at scale has largely limited the use of autotuning to libraries for shared-memory and GPU architectures.
We introduce a framework for \textit{approximate autotuning} that achieves a desired confidence in each algorithm configuration's performance by constructing confidence intervals to describe the performance of individual kernels (subroutines of benchmarked programs).
Once a kernel's performance is deemed sufficiently predictable for a set of inputs, subsequent invocations are avoided and replaced with a predictive model of the execution time.
We then leverage online execution path analysis to coordinate selective kernel execution and propagate each kernel's statistical profile.
This strategy is effective in the presence of frequently-recurring computation and communication kernels, which is characteristic to algorithms in numerical linear algebra. 
We encapsulate this framework as part of a new profiling tool, Critter, that automates kernel execution decisions and propagates statistical profiles along critical paths of execution.
We evaluate performance prediction accuracy obtained by our selective execution methods using state-of-the-art distributed-memory implementations of Cholesky and QR factorization on Stampede2, and demonstrate speed-ups of up to 7.1x with 98$\%$ prediction accuracy.

\end{abstract}


\section{Introduction}\label{intro}
  Distributed-memory schedules of algorithms with limited parallelism exhibit complex trade-offs in synchronization, communication, and computational costs. Online measurements of these costs, the most precise of which model execution time along the critical path, play a significant role in performance prediction, modeling, and analysis. However, analytic costs, and performance modeling in general, cannot alone capture the efficiency of a schedule's underlying computational kernels and communication routines in the presence of an increasingly diverse and complex set of architectures.\par
  A key consequence of recent architectural trends is the emergence of new algorithms in dense linear algebra and other domains that seek to minimize communication and synchronization costs, while balancing work among processors.
  Typically, these algorithms feature a trade-off between various costs (computation, communication, synchronization, memory footprint), which is achieved by a mix of optimizations such as multi-level blocking, alternate scheduling protocols, and processor grid selection~\cite{ballard2014reconstructing}.
The high-dimensional configuration spaces characteristic to these algorithms exacerbate the difficulty of finding an algorithm's optimal configuration. Further, increasing architectural complexity precludes configuration search strategies from easily narrowing the search space to a small set of configurations.
\par
  Autotuning serves as the most precise technique in determining the optimal algorithmic configuration. This technique generates a search space of feasible configurations, which is subsequently pruned as configurations are executed and analyzed. As these configurations often feature competing analytic costs, autotuning is necessary to identify those that best leverage a particular architecture. However, the use of autotuning has been largely limited to shared-memory libraries, as distributed-memory autotuning necessitates consideration of more scales of parallelism, input size, and algorithmic parameters.\par

  Despite the large dimension of an algorithm's configuration space, the number of distinct computation and communication kernels (routines with a particular input size) along individual execution paths is often limited. Performance-limiting algorithms pervasive in scientific computing applications, especially those within graph computations and dense linear algebra, often exhibit frequently-recurring kernel substructure.
  While optimal sampling of configuration spaces has been studied extensively, it is unknown whether autotuning execution time can be further reduced by executing a distributed-memory schedule's kernels selectively. Of equal importance is whether such an approach can attain a desired accuracy in estimating configuration performance.\par


    This work provides a framework for acceleration of autotuning, which is driven by a statistical model of kernel performance.
    This approximate autotuning framework estimates configuration execution time to desired confidence tolerance $\epsilon$ by estimating the execution time of individual kernels along the critical path instead of executing them. 
    A kernel is no longer executed once its sample mean's confidence interval size falls below a threshold. 
    We leverage online execution path analysis both to coordinate selective kernel execution and propagate the performance statistics of distinct kernels along critical paths of execution. 
    Knowing that the number of times a kernel is executed along the critical path is $\alpha$ allows us to assign a sample variance $\sigma^{2}/\alpha$ to a kernel's execution time. 
    The smaller variance reduces the confidence interval necessary to attain desired confidence in the configuration execution time by a factor $\sqrt{\alpha}$.\par

    We encapsulate the proposed framework in an accessible profiling library, Critter. 
    Critter accelerates autotuning by selectively executing kernels and predicting application execution costs. It uses a critical path performance profile of each kernel that is constructed online. 
    In order to limit variability in execution environment, Critter makes consistent selective execution decisions by propagating kernel performance statistics along cartesian processor grids. 

    We address practical challenges for autotuning at scale by presenting the first empirical evidence that selective execution of computation and communication kernels can accelerate automatic performance tuning of distributed-memory dense linear algebra libraries. In addition, we show that guarantees on performance prediction accuracy can be achieved via a priori confidence tolerance specification. Four case studies in autotuning state-of-the-art distributed-memory algorithms for dense Cholesky and QR factorization demonstrate that Critter can accelerate tuning of dense linear algebra programs.
\par

  Our specific contributions are as follows,
  \begin{itemize}
    \item an approximate autotuning methodology that selectively executes kernels using confidence measures to achieve tunable accuracy in algorithm performance prediction,
    \item Critter, an MPI profiling tool for online execution-path analysis that automates selective kernel execution,
    \item empirical evidence that our methodology can accelerate autotuning of state-of-the-art algorithms for dense matrix factorizations; we observe speedups up to 7.1x with $98\%$ accuracy in execution-time prediction accuracy.
  \end{itemize}

\section{Online critical path analysis}\label{schedule_analysis}

  \begin{figure}[t]
    \centering
    \includegraphics[scale=0.2]{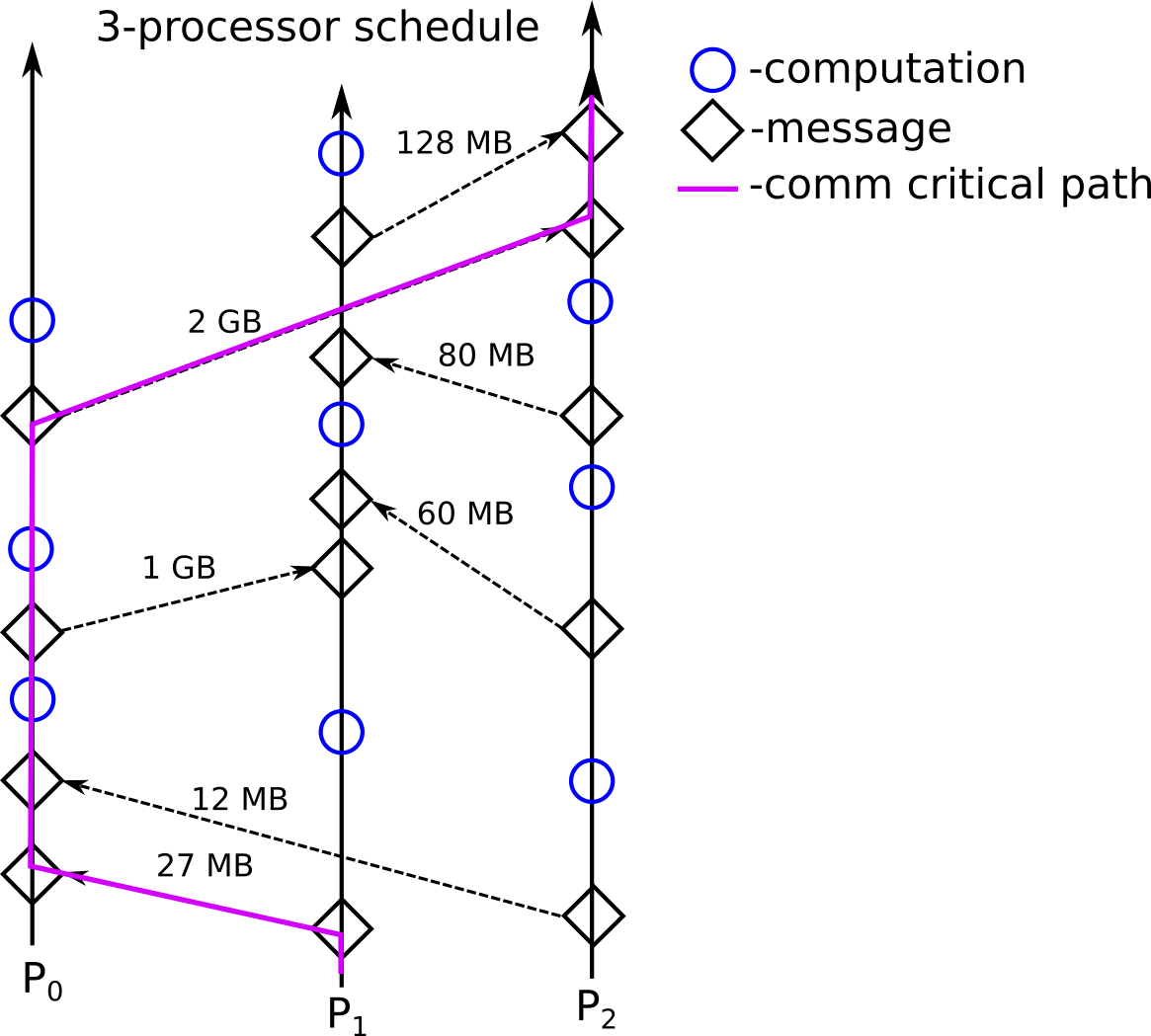}
    \caption{An example of a 3-processor schedule, with an execution path that maximizes communication cost ``comm critical path'', highlighted.}
    \label{fig:cp}
  \end{figure}

  Execution-path analysis models execution time, detects load imbalance, and identifies scaling inefficiencies.
   A key advantage of \textit{online} execution-path analysis is the ability to identify performance bottlenecks at scale. 
   To do so, Critter profiles computation or communication kernels  to construct a profile of the critical path  on the fly. 
  We analyze schedules online to accelerate learning of kernel performance by constructing statistical profiles and facilitating selective execution for each kernel to achieve an accurate overall execution time estimate.

  \subsection{Schedules and cost metrics}
     The execution of an MPI program describes a parallel schedule and its associated execution paths. 
     Directed acyclic graphs (DAGs) $G=(V,E)$ formalize parallel schedules. 
     Edge set $E$ defines the sequences of computation and communication kernels $v\in V$ along which each processor participates (i.e., execution paths).
     A sequence of kernels $\tilde{v}=(v_{1},...,v_{n})$ characterizes an execution path $\tilde{v}$.
     $\tilde{V}$ defines the set of distinct execution paths (all possible paths in $G$).
     Each processor $p_{i}$ participates in a sequence of kernels which itself defines an execution path.\par

    While $|\tilde{V}|$ exhibits an exponential growth rate in the number of communication kernels, a single \textit{critical} path defines a schedule's execution time.
    The critical-path cost of schedule $G=(V,E)$ is defined as $c_{\varphi}(G)=\max_{\pi\in \tilde{V}} \sum_{v\in\pi} \varphi(v)$, where $\varphi : V\rightarrow \mathbb{R}^{+}$ assigns each kernel an execution cost metric.
    Figure \ref{fig:cp} illustrates a three-processor schedule with various execution paths designated by black lines. The execution path that incurs the maximum communication cost may be distinct from the execution path that incurs the maximum execution time.\par

    For a particular MPI program, the set of computational kernels in $V$ can be specified at different granularities. 
    Finer granularity is often desired for offline analysis, and efforts to reduce the overhead in acquiring events along execution paths with program-statement attribution have been studied~\cite{tallent2017representative}. 
    As online path analysis can require substantial computing resources, coarse granularity can reduce complexity in determining allocation of optimization effort across kernels.

  \subsection{Path propagation mechanisms}

    We consider path propagation mechanisms as the additional communication and computation (i.e., additional to that already being done by the application code) necessary to propagate measurement data along execution paths. 
    We use the \textit{pathset} terminology as introduced in \cite{tallent2017representative}. A pathset $\mathcal{P}$ describes a container of paths (chain of dependent or consecutively executed events) or path profiles (statistics/metrics along a particular path). Path profiles store aggregate statistics based on a set of events at varying granularity along a specific path. Pathsets may concatenate those events, although events would be distributed among processors to avoid communication and replication of information.\par

    Path propagation mechanisms for online analysis intercept application communication and propagate measurements along execution paths to participating processors during program execution~\cite{tallent2017representative,10.1145/238020.238024}. 
    For example, the \textit{slack method}~\cite{4154078} filters out paths incurring idle time, while the \textit{longest-path algorithm} isolates paths incurring maximum execution time~\cite{10.1145/238020.238024}.
    Offline profiling mechanisms have overhead relative to online profiling due to increased memory footprint. For example, some mechanisms for offline analysis save performance profiling data (pathset $\mathcal{P}$) to disk intermittently during execution, while others require a backward pass over the communication patterns to identify the root causes of wait states~\cite{bohme2012scalable,hendriks2012reconstructing,4154078,yang1988critical,doi:10.1177/1094342016661865,bohme2016identifying}.\par

    \vspace{0.0cm}

\section{Approximate autotuning framework}\label{approx_autotuning_methodology}

\begin{figure*}
\begin{multicols}{2}
\begin{lstlisting}[mathescape=true,escapeinside={(*}{*)}]
(*\textbf{MPI$\_$Init}*)(argc,argv){
  PMPI_Init(argc,argv);
  // Initialize world communicator as channel
  channel = {}; channel.offset=0; channel.stride[0]=1;
  channel.size[0]=|MPI_COMM_WORLD|; channel.is_maximal=1;
  // Hash id generated purely from (stride,size)
  aggregate_channels[channel.hash] = channel;
}
(*\textbf{MPI$\_$Comm$\_$split}*)(comm,color,key,newcomm){
  PMPI_Comm_split(comm,color,key,newcomm);
  // Calculate (stride,size) of each cartesian dimension
  world_rank = PMPI_Comm_rank(MPI_COMM_WORLD);
  // Initialize array 'ranks' with size |newcomm|
  PMPI_Allgather(world_rank,ranks,newcomm); sort(ranks);
  ~channel = initialize_channel(ranks)~;
  aggregate_channels[channel.hash] = channel;
  // Recursively build aggregate channels
  for (agg in aggregate_channels){
    if (~agg $\notin$ channel~ && ~channel $\notin$ agg~){
      if (~|agg $\cup$ channel|==1~){
        agg.is_maximal = false;
        // Construct new aggregate channel
        PMPI_Allgather(agg.offset,ranks,newcomm);
        sort(ranks); new_hash = agg.hash^channel.hash;
        ~updated_channel = initialize_channel(ranks)~;
        aggregate_channels[new_hash] = updated_channel;}}}
}
(*\textbf{initialize$\_$msg}*)(envelope){
  key,channel = ~generator~(envelope); int_msg.clear();
  if (key $\notin$ $\overline{\mathcal{K}}$) $\overline{\mathcal{K}}$[key]={},$\tilde{\mathcal{K}}$[key]={};
  int_msg.execute = $\tilde{\mathcal{K}}$[key].is_pred;
  int_msg.exec_time = $\mathcal{P}$.exec_time;
  for (i=0; i<$|\tilde{\mathcal{K}}|$; i++){
    int_msg.keys.append($\tilde{\mathcal{K}}$[i].key);
    int_msg.freqs.append($\tilde{\mathcal{K}}$[i].freq);}
  return key,channel,int_msg;
}
// All blocking p2p communications handled similarly
(*\textbf{MPI$\_$Recv}*)(user_msg,envelope){
  key,channel,int_rmsg = initialize_msg(envelope);
  @PMPI_Sendrecv(int_smsg,int_rmsg,envelope)@;
  ~int_rmsg.execute = max(int_smsg.execute,int_rmsg.execute)~;
  if (int_smsg.exec_time > int_rmsg.exec_time){
    $\tilde{\mathcal{K}}$[:].key = int_smsg.keys; $\tilde{\mathcal{K}}$[:].freq = int_smsg.freqs;}
  $\mathcal{P}$.exec_time = max(int_smsg.exec_time,int_rmsg.exec_time);
  if (int_rmsg.execute){
    comm_time = PMPI_Recv(user_msg,envelope);
    ~update_statistics~($\overline{\mathcal{K}}$[key],$\tilde{\mathcal{K}}$[key],comm_time);}
  else comm_time = $\overline{\mathcal{K}}$[key].mean;
  $\mathcal{P}$.exec_time += comm_time;
}
(*\columnbreak*)
// All blocking collective communications handled similarly
(*\textbf{MPI$\_$Bcast}*)(user_msg,envelope){
  key,channel,int_msg = initialize_msg(envelope);
  for (i=0; i<$|\tilde{\mathcal{K}}|$; i++){
    if ($\overline{\mathcal{K}}$[i].is_pred && !$\tilde{\mathcal{K}}$[i].is_pred &&
        channel $\notin$ $\tilde{\mathcal{K}}$[i].agg_channels &&
        $\tilde{\mathcal{K}}$[i].hash ^ channel.hash $\notin$ aggregate_channels){
      save_kernel.append($\overline{\mathcal{K}}$[i],$\tilde{\mathcal{K}}$[i]);
      int_msg.kernel_agg_count++;}}
  PMPI_Allreduce(int_msg,int_gmsg,custom_op,channel);
  $\mathcal{P}$.exec_time = int_gmsg.exec_time;
  ~int_msg.execute = int_gmsg.execute>0~;
  if (int_msg.exec_time < int_gmsg.exec_time){
    $\tilde{\mathcal{K}}$[:].key = int_gmsg.keys; $\tilde{\mathcal{K}}$[:].freq = int_gmsg.freqs;}
  if (int_msg.execute){
    comm_time = PMPI_Bcast(user_msg,envelope);
    ~update_statistics~($\overline{\mathcal{K}}$[key],$\tilde{\mathcal{K}}$[key],comm_time);}
  else comm_time = $\overline{\mathcal{K}}$[key].mean;
  $\mathcal{P}$.exec_time += comm_time;
  ~aggregate_statistics~($\overline{\mathcal{K}}$[key],$\tilde{\mathcal{K}}$[key]);
}
// All nonblocking p2p communications handled similarly
(*\textbf{MPI$\_$Isend}*)(user_msg,envelope,request){
  key,channel,int_smsg = initialize_msg(envelope);
  @PMPI_Bsend(int_smsg,envelope)@;
  if (int_smsg.execute){
    PMPI_Isend(user_msg,envelope,request);}
  else @request = request_generator()@;
  @int_request = request_generator()@;
  nonblocking_dict[request] = (key,channel,int_smsg,int_request);
}
// All nonblocking collective communications handled similarly
(*\textbf{MPI$\_$Iscatter}*)(user_msg,envelope,request){
  key,channel,int_msg = initialize_msg(envelope);
  PMPI_Iallreduce(int_msg,int_gmsg,envelope,request);
  if (int_msg.execute){
    PMPI_Iscatter(user_msg,envelope,request);}
  else @request = request_generator()@;
  @int_request = request_generator()@;
  nonblocking_dict[request] = (key,channel,int_msg);
}
// All outstanding request completion ops handled similarly
(*\textbf{MPI$\_$Wait}*)(request,status){
  key,channel,int_msg,int_request = nonblocking_dict[request];
  if (int_msg.execute){
    comm_time = PMPI_Wait(request,status);
    ~Update_statistics~($\overline{\mathcal{K}}$[key],$\tilde{\mathcal{K}}$[key],comm_time);}
  else comm_time = $\overline{\mathcal{K}}$[key].mean;
  @PMPI_Wait(int_request,status)@;
  if (int_msg.exec_time > $\mathcal{P}$.exec_time){
    $\tilde{\mathcal{K}}$[:].key = int_msg.keys; $\tilde{\mathcal{K}}$[:].freq = int_msg.freqs;}
  $\mathcal{P}$.exec_time = max($\mathcal{P}$.exec_time,int_msg.exec_time);
}
\end{lstlisting}
\end{multicols}
\caption{Description of our framework's mechanism, which propagates statistical profiles upon interception of various communication primitives. \color{blue}Blue text \color{black} signifies path propagation logic generic to online critical-path analysis that can be modified to reflect various protocols. \color{red}Red text \color{black} signifies path propagation logic specific to calculating kernel performance statistics.}
\label{path_prop_alg}
\end{figure*}

  Autotuning consists of benchmarking an algorithm over a space of tuning parameters (e.g., block sizes) and test problems to determine an optimal configuration. The technique is widely used in the design of modern high-performance computing software~\cite{balaprakash2018autotuning}, but can be prohibitively expensive when tuning algorithms from dense linear algebra or other application domains at scale. We now specify our framework for acceleration of autotuning, which is driven by a statistical model of kernel (routine with a particular input size) performance.\par

\subsection{Statistical characterization of execution paths}

  We introduce an approach to autotuning that estimates a configuration's execution time $c_{\varphi}$ to a fixed confidence level $\epsilon$. We formalize this approach under the unifying assumption that a metric measurement of each kernel's execution time follows a distribution with finite mean $\mu$ and variance $\sigma^{2}$. 
  Specifically, we assume that an executed kernel's measured performance $\varphi(V)$ is a random variable $X$ drawn from a distribution that is the same for all kernels with a given signature (i.e., program function for a given input size). 

  To estimate $c_{\varphi}$ to a fixed confidence level, we construct a confidence interval for the performance of each set of executed kernels with the same signature scheduled along any sub-critical path in $G$. A sub-critical path is the critical path of some subgraph of $G$ constructed from the start of execution to some intermediate kernel.
In particular, for each set of $k$ kernels $W=\{w_{1},\ldots, w_{k}\}$ with the same signature (hence modeled by the same random variable $X$) appearing along the same sub-critical path, standard single-pass algorithms are used construct kernel performance models. 
  We use the execution time sample mean $\Expe[X] \approx \bar{w} = \frac{1}{k}\sum_{w_i\in W}\varphi(w_{i})$ and estimate the variance for the random variable $T$ representing the combined time for these kernels $\Var[T] =\frac{1}{\sqrt k} \Var[X] \approx \frac{1}{k^{3/2}} \sum_{i\in W}(\bar{w} - w_{i})^2$ during program execution. Knowledge of this sample count $k$ decreases the confidence interval size for $\bar{w}$ by a factor $\sqrt{k}$, and thus can significantly reduce the number of kernel executions necessary to estimate $c_{\varphi}$ in our framework.


      Our autotuning framework predicts a configuration's execution time by collecting samples to construct models of the execution time of individual kernels until each is predictable to confidence tolerance $\epsilon$ (i.e., the confidence interval size divided by sample mean, denoted by $\tilde{\epsilon}$, satisfies $\tilde{\epsilon}\leq \epsilon$). 
      Its statistical characterization of kernel performance along sub-critical paths offers a trade-off between prediction accuracy of a configuration's performance and the speed at which that accuracy is attained. 
      In particular, prediction accuracy can be systematically improved by incrementally decreasing the confidence tolerance $\epsilon$ of each kernel's sample mean $\bar{w}$ until kernel execution is never avoided and accuracy is maximal.\par

\subsection{Propagation mechanism for selective execution policies}

  We present a path propagation mechanism that enables our approximate autotuning framework by coordinating the selective execution of kernels (routines with a particular input size). 
  This mechanism constructs a critical-path cost model of each configuration during program execution using kernel measurements sampled along sub-critical execution paths. 
  It leverages the longest-path algorithm~\cite{10.1145/238020.238024} to propagate a configuration's execution costs and the performance statistics (i.e., the sample mean, sample variance, and execution count) of its kernels. 
  Each processor owns two distinct sets, $\overline{\mathcal{K}}$ and $\tilde{\mathcal{K}}$, that store performance statistics for each locally-executed kernel and each kernel executed along its current sub-critical path, respectively. 
  The information in these sets is used to determine whether a kernel's execution time is sufficiently predictable.

  Kernel execution decisions in our framework take into account not only the performance statistics of a local processor's kernels, but those of remote processors as well. 
  We describe multiple kernel execution policies that use online path propagation to enforce constraints on kernel execution decisions.
  Each policy propagates $\tilde{\mathcal{K}}$ among subsets of processors along sub-critical execution paths. 
  The additional profiling costs attributed to each policy depend on the degree to which kernel performance statistics on remote processors affect a processor's kernel execution decisions. 
  By default, processors determine whether to execute computational kernels independently, while communication kernel execution is skipped only if a sufficiently-large number of processors within the sub-communicator associated with the kernel deem its execution time to be sufficiently predictable.

  Kernel performance can be sensitive to changes in available hardware resources, e.g., due to contention among processes on the same node for memory bandwidth and network resources. 
  As such, inconsistent execution policies can cause variability in kernel timings. 
  To circumvent this, we also consider synchronization of kernel execution decisions among all processors. 
  To accomplish this without interfering with application execution, we propagate kernel performance statistics among dimensions of a cartesian processor grid. 

  We achieve this by recursively building \textit{aggregates}: subsets of channels that serve as a basis for a cartesian processor grid.
  An aggregate channel defining the complete processor grid is constructed when \texttt{MPI$\_$Init} is intercepted in Figure \ref{path_prop_alg}.
  Subsequent aggregates are constructed when \texttt{MPI$\_$Comm$\_$split} is intercepted by using internal communication to attain processor offsets, spans, and sub-communicator sizes.
  Propagation of kernel performance statistics along subsequent channels is permitted if the stride and size of the new communicator, when combined with those along which the sample has been previously propagated, yields a cartesian processor grid.
  Once a kernel's performance statistics have been propagated so as to establish policy agreement among all processors, its execution can be switched off.
  We remark that this infrastructure for selective propagation of performance statistics across distinct cartesian communication channels can also be used to prevent sampling bias that would arise when aggregating measurement samples along overlapping partitions of a processor grid.
  We do not evaluate its use for that purpose in this work.\par 

\section{Approximate autotuning instrumentation}\label{critter_sec}
  The profiling overhead necessary to quantify confidence along critical execution paths to a desired confidence tolerance $\epsilon$ can be amortized by leveraging overlapping kernel substructure. 
  To demonstrate practicality of the methods introduced in Section \ref{approx_autotuning_methodology}, we have developed Critter\footnote{\url{https://github.com/huttered40/critter}}, a lightweight profiling tool that accelerates tuning of MPI distributed-memory (dense) linear algebra software using online execution-path analysis.\par

  \subsection{Instrumentation}
  Critter implements the path propagation mechanism described in Section \ref{approx_autotuning_methodology} for collective and point-to-point communications via both blocking and nonblocking protocols. 
  C++ programs use Critter by including a header file and inserting start/stop calls to the Critter profiler at the beginning and end of execution.
  Critter intercepts and selectively executes MPI, BLAS, and LAPACK routines automatically, and allows library developers to selectively execute loop nests and other structures by inserting preprocessor directives. 
  Its instrumentation is described in Figure \ref{path_prop_alg} for the following MPI routines: \texttt{Recv}, \texttt{Isend}, \texttt{Bcast}, \texttt{Iscatter}, and \texttt{Wait}.\par

  Upon interception of each communication kernel, a \textit{generator} method determines the signature from the message envelope, an internal message is constructed with kernel performance statistics from $\tilde{\mathcal{K}}$, and an execution decision propagated. 
  A reduction is also performed among participating processors, which serves to differentiate communication time from idle time and incrementally construct a critical path performance model. 
  After the reduction of kernel set $\tilde{\mathcal{K}}$, the user communication kernel is selectively executed, its statistics are modified by invoking \textit{update$\_$statistics}, and the pathset $\mathcal{P}$ accumulates the kernel's execution time.
  Blocking collectives can additionally invoke \textit{aggregate$\_$statistics} to propagate a kernel's performance statistics across the sub-communicator if its performance is sufficiently predictable.\par 

  Although not provided explicitly in Figure \ref{path_prop_alg}, Critter can also profile MPI routines with other synchronization mechanisms. 
  Our default (non-synchronized) selective execution protocol is used to handle nonblocking communication routines. 

  \subsection{Methods for selective kernel execution}
  Critter provides multiple methods to calculate kernel performance statistics described in Section \ref{approx_autotuning_methodology}. 
  Each method constructs the mean and variance of a kernel's performance distribution independently among processors (i.e., using measurements of locally-executed kernels), rather than using only those kernel executions invoked along the current sub-critical execution path as described in Section \ref{approx_autotuning_methodology}. 
  However, each kernel's confidence interval size is determined only when its execution count (i.e., the number of times it's executed on some execution path) has been updated to reflect the current sub-critical execution path.\par

  We first consider \textit{eager propagation}, a method that skips kernel execution once a single processor deems that kernel sufficiently predictable and its corresponding performance statistics has been propagated across all processors. This method leverages the aggregate channel infrastructure introduced in the previous section and is intended for bulk-synchronous algorithms. 
  All subsequent methods adopt an execution policy that allows processors to determine computation kernel decisions independently, and avoids execution of communication kernels only if all processors within the corresponding kernel's sub-communicator deem its performance sufficiently predictable. 
  These methods differ in how they propagate kernel execution counts to determine the desired prediction confidence level. 

  The \textit{online propagation} method performs online propagation of kernel execution counts between processors. 
  Each processor updates confidence interval sizes on-the-fly to reflect a kernel's execution count along the current sub-critical path.
  The \textit{local propagation} method 
  does not propagate information among processors, collecting only local kernel statistics. 
  \textit{A priori propagation} forgoes online propagation by using an initial offline iteration to allow immediate application of critical-path kernel execution counts in determining confidence interval sizes, but performs online propagation of kernel statistics to construct a model. 
  The most conservative method, \textit{conditional execution}, also forgoes online propagation and does not use kernel execution counts to reduce the confidence tolerance.

\section{Application to Dense Linear Algebra}\label{analysis_sec}

Significant effort goes into tuning dense linear algebra kernels to achieve maximal performance on a single node and across full supercomputers (e.g., to tune LINPACK~\cite{dongarra1979linpack,dongarra2003linpack}). 
We apply Critter to autotune two dense matrix factorization algorithms: Cholesky and QR factorization.
These algorithms enable the solution of linear systems of equations and linear least squares problems.

We evaluate Critter and its underlying methodology using four distinct state-of-the-art library implementations of QR and Cholesky.
The chosen libraries implement both bulk-synchronous algorithms (i.e., algorithms decomposed into alternating synchronous steps of asynchronous communication and computation) and task-based algorithms in which computation and communication overlap. 

For Cholesky, we first consider Capital's recursive bulk-synchronous algorithm on a partially-replicated cyclic matrix distribution over a 3D processor grid~\cite{HutterCapital2019}, which serves as a key subroutine in a recent communication-avoiding QR factorization algorithm~\cite{hutter2019communication}. We next consider Slate's Cholesky routine, which uses an iterative algorithm on a block-cyclic matrix distribution that performs task-based scheduling over a 2D processor grid~\cite{gates2019slate}. 

For QR, we study two iterative algorithms on block-cyclic matrix distributions across 2D processor grids. Slate's QR routine utilizes task-based scheduling~\cite{kurzak2019least} while CANDMC uses a pipelined bulk synchronous algorithm~\cite{SolomonikCANDMC}.
  We provide costs in the bulk-synchronous-parallel (BSP) model for some of the Cholesky and QR algorithms.
  \subsection{Cholesky factorization}
    Given a symmetric positive definite matrix $A$, the Cholesky factorization computes a lower-triangular matrix $L$ such that $A=LL^{T}$.
    Capital's communication-efficient algorithm can be obtained via recursive application of Cholesky~\cite{Tiskin2002} from 
      \begin{align*}
        \begin{bmatrix}
          A_{11} & A_{21}^{T}\\
          A_{21} & A_{22}
        \end{bmatrix}&=
        \begin{bmatrix}
          L_{11} &\\
          L_{21} & L_{22}
        \end{bmatrix}
        \begin{bmatrix}
          L^{T}_{11} & L^{T}_{21}\\
          & L^{T}_{22}
        \end{bmatrix},\\
        \begin{bmatrix}
          I &\\
           & I
        \end{bmatrix}&=
        \begin{bmatrix}
          L_{11} &\\
          L_{21} & L_{22}
        \end{bmatrix}
        \begin{bmatrix}
          L^{-1}_{11} &\\
          S_{21} & L^{-1}_{22}
        \end{bmatrix}.
      \end{align*}
    The base case is solved with sequential BLAS routines once the sub-problem is sufficiently small (i.e., dimension is below some block size).
    Aside from recursive calls, the algorithm uses products of triangular and square matrices ($L_{21} \gets A_{21}L_{11}^{-T}$ and $S_{21} \gets -L^{-1}_{22}L_{21}L^{-1}_{11}$) as well as the symmetric rank-$k$ update ($A_{22}-L_{21}L^{T}_{21}$).

    Both types of matrix--matrix products are executed with minimal communication using a 3D processor grid~\cite{matmul3d,snirmatmul,Johnsson:1993:MCT:176639.176642,dekel:657}, with broadcasts along two dimensions of the processor grid, and a reduction along the third.
    While communication-efficient, using a 3D grid also entails additional memory footprint; each of $p^{1/3}$ copies of $A$, $L$, and $L^{-1}$ is distributed cyclically among $p^{2/3}$ processors.
    For a matrix of dimension $n$, given base-case block size $b$, with any base case strategy, the BSP cost of Capital's Cholesky algorithm is~\cite{HutterCapital2019}
    \[\Theta\Big(\alpha \cdot n/b + \beta \cdot (n^2/p^{2/3} + nb) + \gamma \cdot (n^3/p + nb^2)\Big).\]
    The first term (latency cost, i.e., number of super-steps) is minimized by a large block size, while the latter two terms (communication and computation cost) are minimized by a small block size.
    Consequently, there is a non-trivial trade-off among costs that makes it difficult to pick the most performant block size a priori. 

    \begin{figure*}[t]
      \centering
      \begin{subfigure}{0.24\textwidth}
        \centering
        \includegraphics[scale=.65]{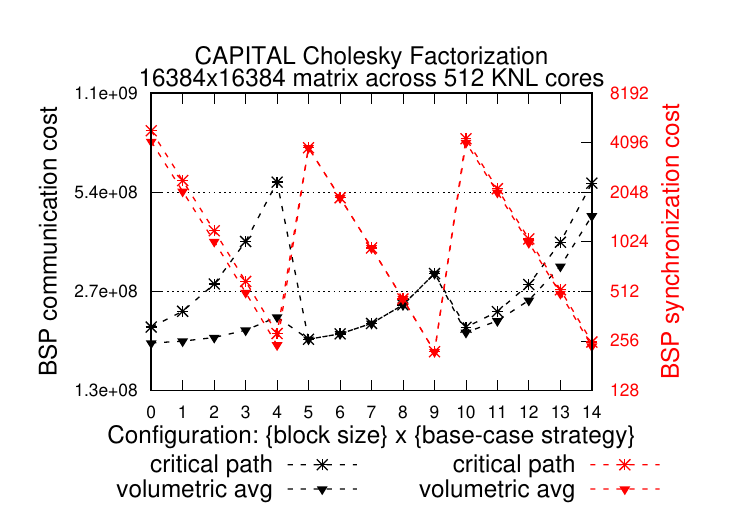}
        \caption{}
        \label{fig:capital_cholesky_comm_synch_vary}
      \end{subfigure}
      \begin{subfigure}{0.24\textwidth}
        \centering
        \includegraphics[scale=.65]{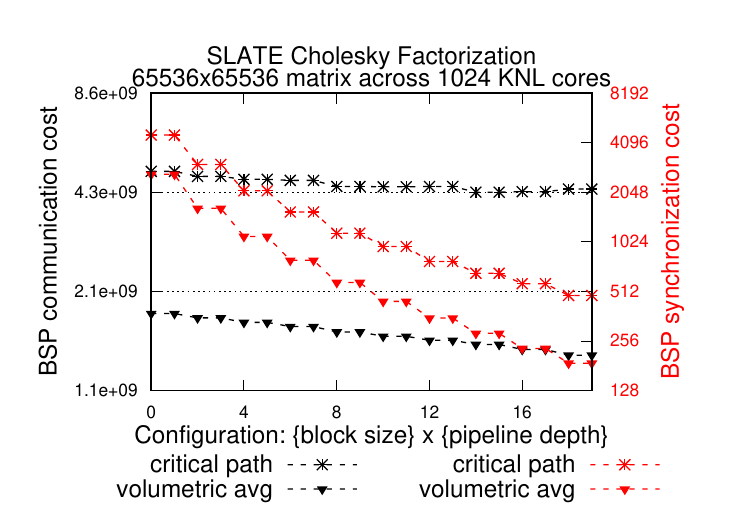}
        \caption{}
        \label{fig:slate_cholesky_comm_synch_vary}
      \end{subfigure}
      \begin{subfigure}{0.24\textwidth}
        \centering
        \includegraphics[scale=.65]{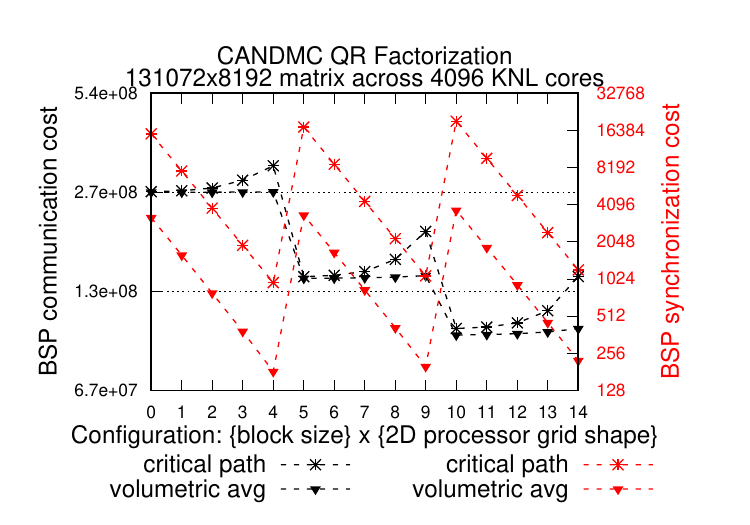}
        \caption{}
        \label{fig:candmc_qr_comm_synch_vary}
      \end{subfigure}
      \begin{subfigure}{0.24\textwidth}
        \centering
        \includegraphics[scale=.65]{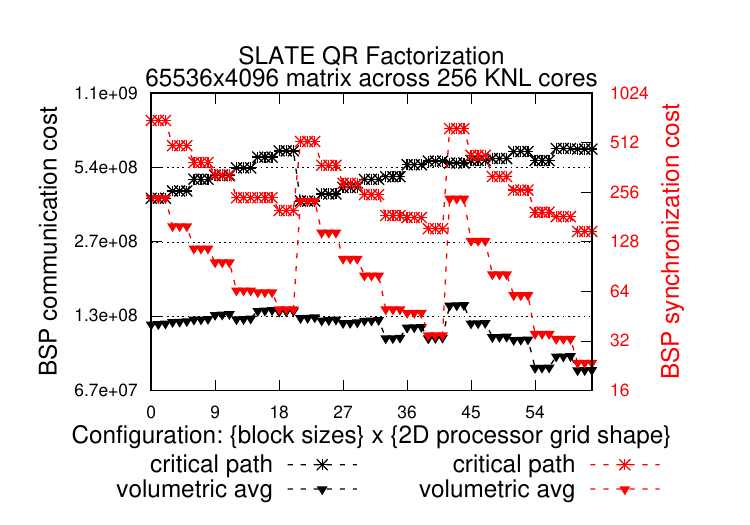}
        \caption{}
        \label{fig:slate_qr_comm_synch_vary}
      \end{subfigure}
      \begin{subfigure}{0.24\textwidth}
        \centering
        \includegraphics[scale=.65]{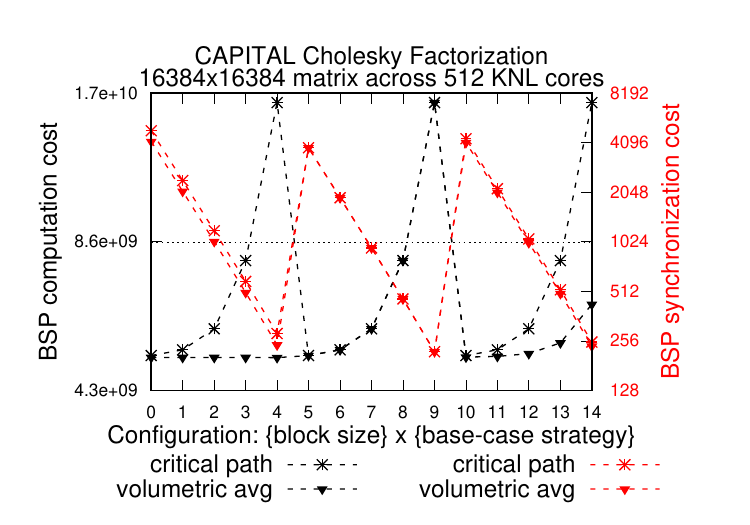}
        \caption{}
        \label{fig:capital_cholesky_comp_synch_vary}
      \end{subfigure}
      \begin{subfigure}{0.24\textwidth}
        \centering
        \includegraphics[scale=.65]{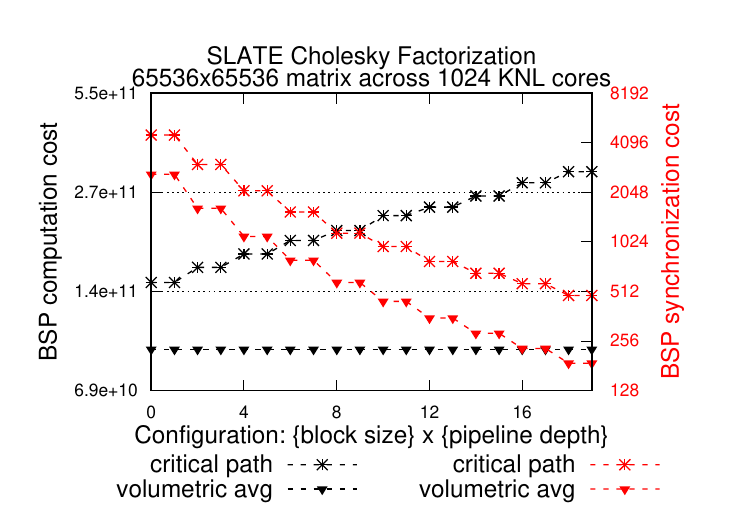}
        \caption{}
        \label{fig:slate_cholesky_comp_synch_vary}
      \end{subfigure}
      \begin{subfigure}{0.24\textwidth}
        \centering
        \includegraphics[scale=.65]{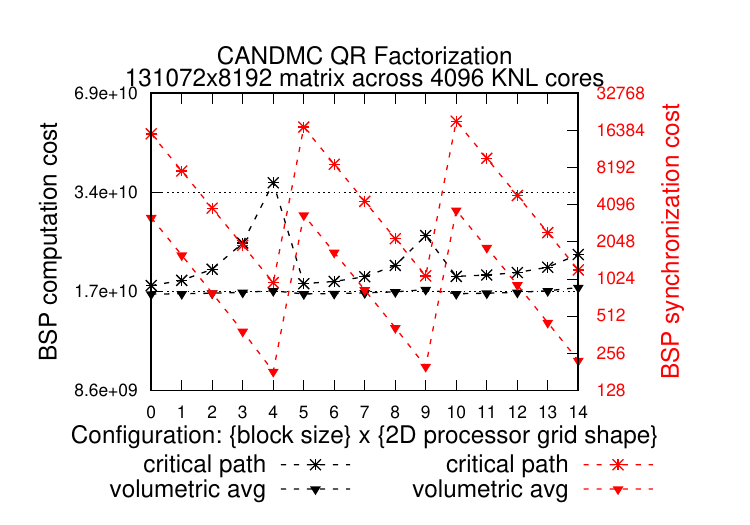}
        \caption{}
        \label{fig:candmc_qr_comp_synch_vary}
      \end{subfigure}
      \begin{subfigure}{0.24\textwidth}
        \centering
        \includegraphics[scale=.65]{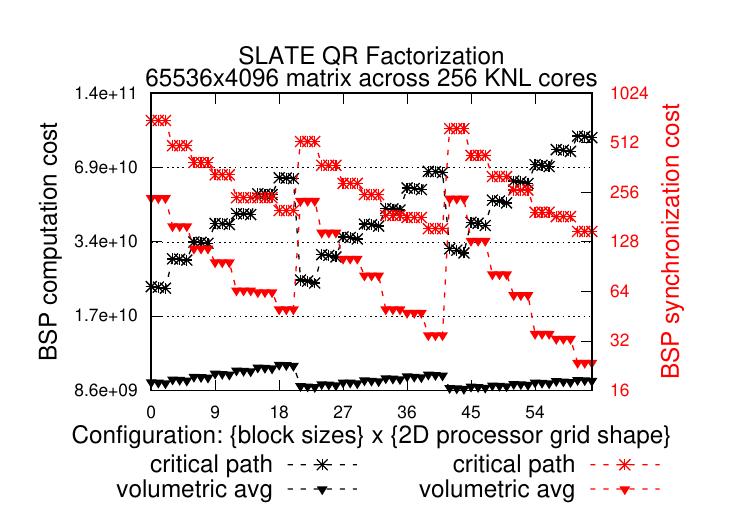}
        \caption{}
        \label{fig:slate_qr_comp_synch_vary}
      \end{subfigure}
      \begin{subfigure}{0.24\textwidth}
        \centering
        \includegraphics[scale=.6]{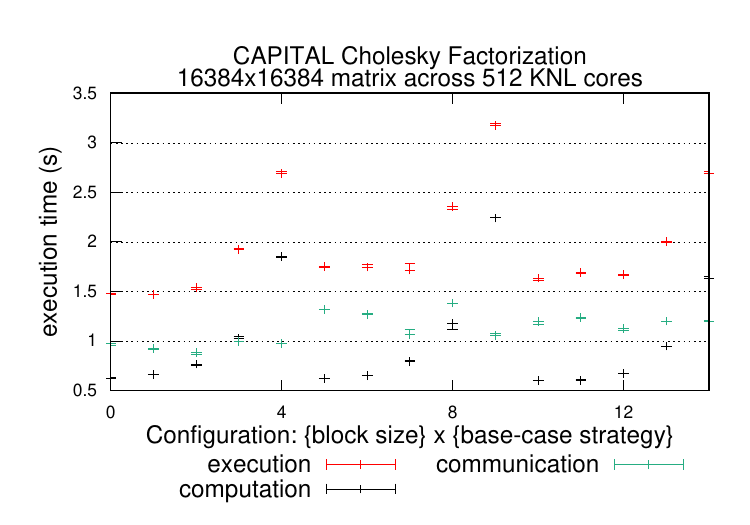}
        \caption{}
        \label{fig:capital_cholesky_time_vary}
      \end{subfigure}
      \begin{subfigure}{0.24\textwidth}
        \centering
        \includegraphics[scale=.6]{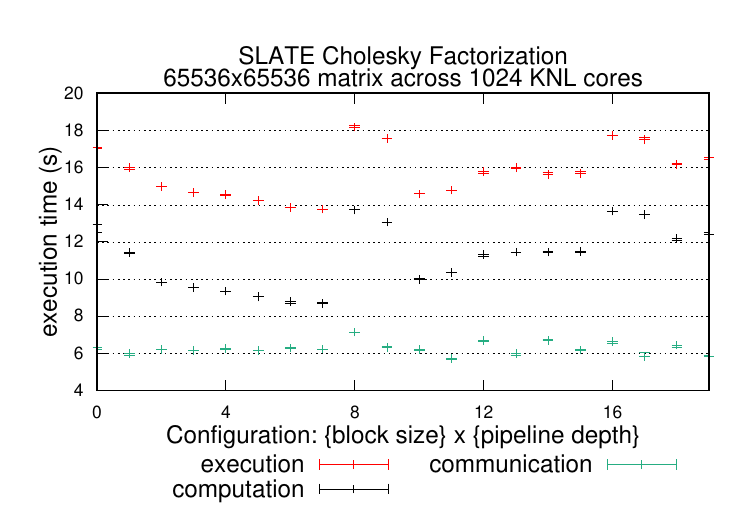}
        \caption{}
        \label{fig:slate_cholesky_time_vary}
      \end{subfigure}
      \begin{subfigure}{0.24\textwidth}
        \centering
        \includegraphics[scale=.6]{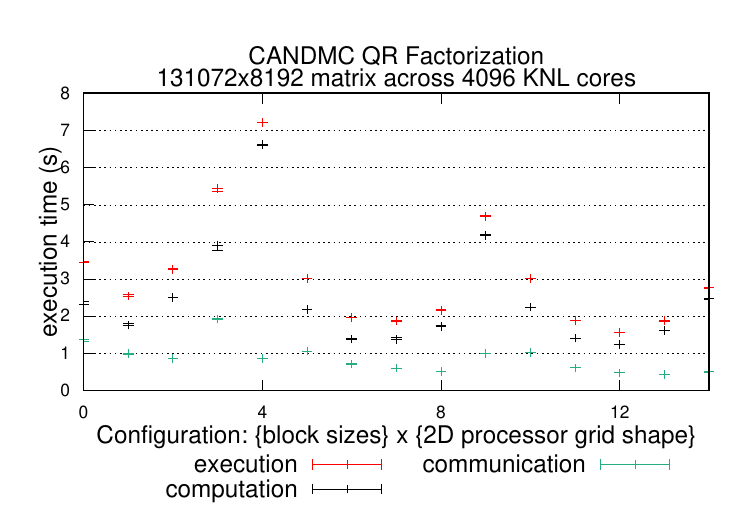}
        \caption{}
        \label{fig:candmc_qr_time_vary}
      \end{subfigure}
      \begin{subfigure}{0.24\textwidth}
        \centering
        \includegraphics[scale=.6]{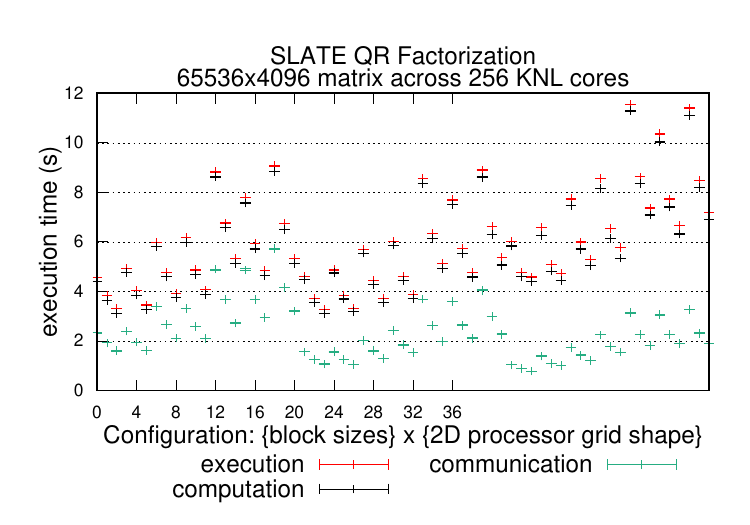}
        \caption{}
        \label{fig:slate_qr_time_vary}
      \end{subfigure}
      \caption[]{Figures \ref{fig:capital_cholesky_comm_synch_vary}, \ref{fig:slate_cholesky_comm_synch_vary}, \ref{fig:candmc_qr_comm_synch_vary}, and \ref{fig:slate_qr_comm_synch_vary} illustrate trade-offs in BSP communication and synchronization, while Figures \ref{fig:capital_cholesky_comp_synch_vary}, \ref{fig:slate_cholesky_comp_synch_vary}, \ref{fig:candmc_qr_comp_synch_vary}, and \ref{fig:slate_qr_comp_synch_vary} illustrate trade-offs in BSP computation and synchronization. The corresponding critical-path execution times for each configuration are given in Figures \ref{fig:capital_cholesky_time_vary}, \ref{fig:slate_cholesky_time_vary}, \ref{fig:candmc_qr_time_vary}, and \ref{fig:slate_qr_time_vary}.}
      \label{fig:vary}
    \end{figure*}

    Aside from choosing the block size that determines the size of the base-case problem, there is also a choice of how to compute base-case problems in a distributed setting.
    We consider three strategies:
    \begin{enumerate}
      \item gather the base-case matrix onto one process within a single processor grid layer, compute its factorization, scatter it across the processor grid layer, and broadcast it along the depth of the processor grid,
      \item all-gather the base-case matrix within each processor grid layer and compute its factorization redundantly,
      \item all-gather the base-case matrix within a single processor grid layer, compute its factorization redundantly across that layer, and broadcast it along the depth of the processor grid.
    \end{enumerate}\par

    Slate's synchronization-efficient algorithm instead immediately partitions the matrix into tiles of some tunable size across a 2D processor grid. Each tile maintains a predecessor list of tasks (i.e., triangular solve $L_{21}L^{T}_{11} = A_{21}$ and the symmetric rank-$k$ update $A_{22}-L_{21}L^{T}_{21}$) that must be completed prior to its own execution. 
    Slate employs look-ahead pipelining with tunable depth, an optimization that seeks to maximize the computation concurrency and thus reduce computation along the critical-path. Its task-based scheduling protocols are implemented using nonblocking point-to-point communication, which aims to further reduce synchronization overheads in practice.\par


  \subsection{QR factorization}
    The CANDMC algorithm for QR factorization~\cite{Demmel:2012:CPS:2340316.2340324,ballard2014reconstructing} successively factorizes panels of the matrix, $A = \begin{bmatrix} A_1 & \cdots & A_r \end{bmatrix}$.
    To factorize $A=QR$ where $Q$ is orthogonal and $R$ is rectangular, the compact Householder representation $Q=I-YTY^T$ is utilized, where $T$ is upper triangular and $Y$ is unit-diagonal and lower-trapezoidal.
    Factorizing a panel of $A$, e.g., $A_1=Q_1R_1$, can be done by a variety of algorithms including TSQR~\cite{Demmel:2012:CPS:2340316.2340324} and Cholesky-QR2~\cite{fukaya2014choleskyqr2}. 
    Reconstruction of the Householder representation, then yields the first panel $Y_1$ of $Y$ by LU factorization of a matrix derived from $Q_1$~\cite{Yamamoto12-talk,ballard2014reconstructing}.
    Given $Y_1$, the algorithm updates other panels of $A$ (these are referred to as the trailing matrix) to be $(I-Y_1T_{11}Y_1^T)^TA_i$ for $i=2,\ldots,r$ and continues on to the factorization of $A_2$.

    CANDMC implements these steps in parallel via a 2D block-cyclic distribution of the matrix data. 
    CANDMC's look-ahead pipelining mechanism ~\cite{1885-40739,kurzak2006implementing} enables processors to work on trailing matrix updates if they are not in the processor grid column that participates in factorizing the panel of the matrix. 
    This optimization aims to reduce the computational cost along the critical path rather than the per-processor workload.\par
    For an $m\times n$ matrix $A$ with $m\geq n$ distributed on a $p_r\times p_c$ processor grid with block size $b\leq \min(m/p_r,n/p_c)$, the CANDMC implementation has the following BSP cost~\cite{ballard2014reconstructing},
    \begin{align*}
    \Theta\Big(&\alpha\cdot n/b + \beta\cdot (mn/p_r+n^2/p_c + nb) \\
+ &\gamma \cdot (mn^2/p + nb^2 + mnb/p_r+n^2b/p_c)\Big).
    \end{align*}\par
    As reflected by the trade-offs in its BSP cost terms, the performance of the QR algorithm is highly sensitive to the choice of block size ($b$) and processor grid ($p_r$, $p_c$).
    The cost analysis allows us to understand the scaling of the best choice of processor grid relative to the dimensions, namely $m/p_r=\Theta(n/p_c)$.
    However, in practice, constant factors associated with these costs and overhead in realistic execution of implementations makes a range of block sizes and processor grids viable candidates for each choice of $m,n,p$.
    Autotuning is often the most effective way of finding the fastest implementation in a region of viable parameters.

    Slate implements a different variant of Householder QR, also leveraging a block-cyclic distribution on a 2D processor grid.
    Like for Cholesky, Slate's algorithm
    for QR makes use of task-based scheduling and and pipelining mechanisms. 
    Additionally, its panel factorization is internally blocked (parameter $w$) to increase thread concurrency.


  \subsection{Schedule configurations}
    We tune Capital's Cholesky algorithm across 15 distinct configurations. Each configuration $v\in [0,14]$ factors a $16384\times 16384$ matrix using $512$ KNL cores, with block size $b=128\cdot 2^{v\%5}$ and base case strategy $\lceil (v+1)/5\rceil$. We tune Slate's \texttt{potrf} routine across 20 distinct configurations. Each configuration $v\in [0,19]$ factors a $65536\times 65536$ matrix using $1024$ KNL cores, with pipeline depth $v\%2$ and tile size $256+64\cdot \lfloor v/2\rfloor$. Excessive computational overhead precluded evaluation of smaller tile sizes.\par
    We tune CANDMC's pipelined 2D QR schedule across 15 distinct configurations. Each configuration $v\in [0,14]$ factors a $131072\times 8192$ matrix using $4096$ KNL cores, with block size $b=8\cdot 2^{v\%5}$ and processor grid dimensions $p_{r}\times p_{c}: 64\cdot 2^{\lfloor v/5\rfloor}\times \lfloor 64/2^{\lfloor v/5\rfloor}\rfloor$. We tune Slate's \texttt{geqrf} routine across 63 distinct configurations. Each configuration $v\in [0,62]$ factors a $65536\times 4096$ matrix using $256$ KNL cores, with smaller panel width ($w$) $8\cdot 2^{v\%3}$, panel width $256+64\cdot \lfloor v/3\rfloor \%7$, and processor grid dimensions $p_{r}\times p_{c}: 64/2^{\lfloor v/21\rfloor}\times 4\cdot 2^{\lfloor v/21\rfloor}$.\par
    These configurations are chosen to exemplify the challenge of choosing parameters to minimize execution time for a particular matrix factorization and problem size. Load imbalance and variations in computational concurrency captured by volumetric and critical-path measurements further exacerbate this challenge. Figures \ref{fig:capital_cholesky_comm_synch_vary}, \ref{fig:slate_cholesky_comm_synch_vary}, \ref{fig:candmc_qr_comm_synch_vary}, and \ref{fig:slate_qr_comm_synch_vary} illustrate cost trade-offs between communication and synchronization in the BSP model. Figures \ref{fig:capital_cholesky_comp_synch_vary}, \ref{fig:slate_cholesky_comp_synch_vary}, \ref{fig:candmc_qr_comp_synch_vary}, and \ref{fig:slate_qr_comp_synch_vary} illustrate trade-offs between computation and synchronization in the BSP model. The corresponding execution times for each configuration are given in Figures \ref{fig:capital_cholesky_time_vary}, \ref{fig:slate_cholesky_time_vary}, \ref{fig:candmc_qr_time_vary}, and \ref{fig:slate_qr_time_vary}.\par

  \subsection{Kernel configurations}
    Like all standard dense matrix factorization algorithms, the Cholesky and QR algorithms we evaluate invoke basic linear algebra subroutines (BLAS)~\cite{lawson1979basic}, LAPACK~\cite{LAPACK} functions, and MPI~\cite{Gropp:1994:UMP:207387} collective and point-to-point communication routines. We leverage Critter's ability to define kernels from arbitrary segments of code in intercepting block-to-cyclic data distribution kernels in Capital's Cholesky algorithm. Otherwise, we limit kernel interception to BLAS, LAPACK, and MPI routines in this work.\par
    Critter parameterizes computational kernels on matrix dimensions and other BLAS parameters such as transposition of individual matrices.
    Critter parameterizes communication kernels on message size as well as the MPI subcommunicator size and stride relative to the world communicator. Point-to-point communication configurations are treated as equivalent to size-2 sub-communicators.
    This description suffices for any communicator along a fiber or across a slice of a processor grid, as needed for dense linear algebra algorithms.\par

    Both distributed-memory Cholesky factorization algorithms utilize the LAPACK Cholesky factorization routine (\texttt{potrf}) as well as BLAS routines for general matrix-matrix products (\texttt{gemm}), symmetric rank-k updates (\texttt{syrk}), and triangular-system solves (\texttt{trsm}). Capital additionally uses the BLAS routine for triangular matrix-matrix products (\texttt{trmm}) and the LAPACK routine for triangular matrix inversion (\texttt{trtri}). Slate's Cholesky algorithm utilizes MPI routines for \texttt{isend} and \texttt{recv} kernels, while Capital utilizes MPI routines for \texttt{bcast}, \texttt{allreduce}, \texttt{reduce}, \texttt{allgather}, \texttt{scatter}, and \texttt{gather}.\par

    Both QR algorithms utilize \texttt{gemm} and LAPACK routines for blocked QR factorization of triangular-pentagonal matrices (\texttt{tpqrt}) and subsequent orthogonal transformations (\texttt{tpmqrt}). Slate's QR implementation additionally uses \texttt{trmm}, while CANDMC additionally uses \texttt{trsm}, \texttt{trtri}, and LAPACK routines for QR factorization of general matrices (\texttt{geqrf}) and subsequent orthogonal transformations (\texttt{ormqr}). CANDMC utilizes MPI routines for \texttt{bcast}, \texttt{allreduce}, \texttt{send} and \texttt{recv}. Slate uses \texttt{isend}, \texttt{send}, and \texttt{recv}. We do not consider selective execution of BLAS-2 kernels invoked by Slate's QR implementation.\par


    \begin{figure*}[htbp]
      \centering
      \begin{subfigure}{0.24\textwidth}
        \centering
        \includegraphics[scale=.6]{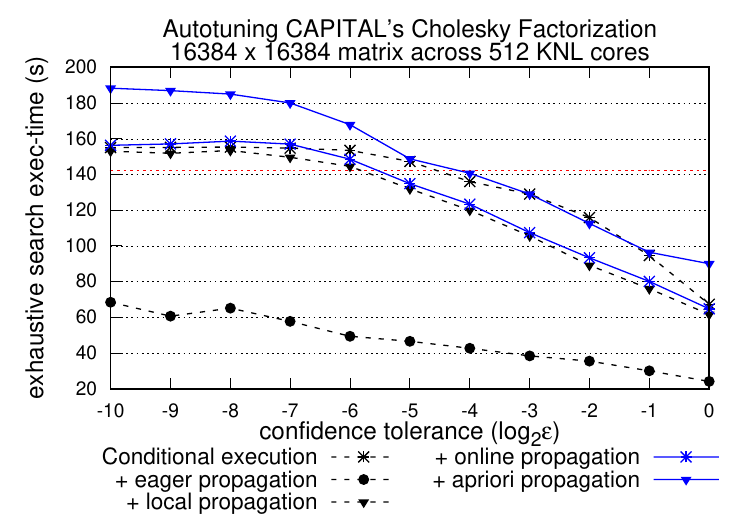}
        \caption{}
        \label{fig:capital_cholesky_tt}
      \end{subfigure}
      \begin{subfigure}{0.24\textwidth}
        \centering
        \includegraphics[scale=.6]{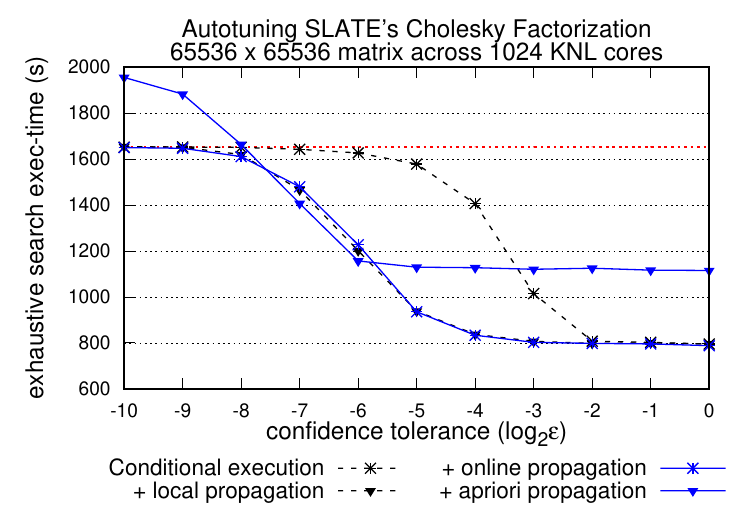}
        \caption{}
        \label{fig:slate_cholesky_tt}
      \end{subfigure}
      \begin{subfigure}{0.24\textwidth}
        \centering
        \includegraphics[scale=.6]{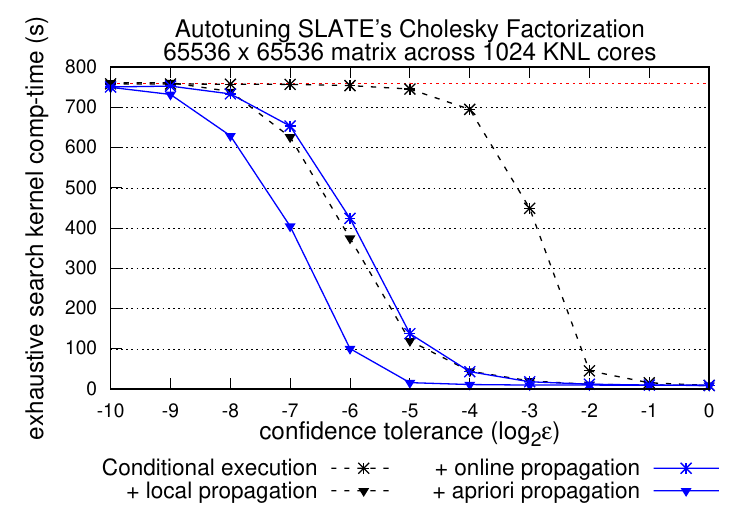}
        \caption{}
        \label{fig:slate_cholesky_comp_tt}
      \end{subfigure}
      \begin{subfigure}{0.24\textwidth}
        \centering
        \includegraphics[scale=.6]{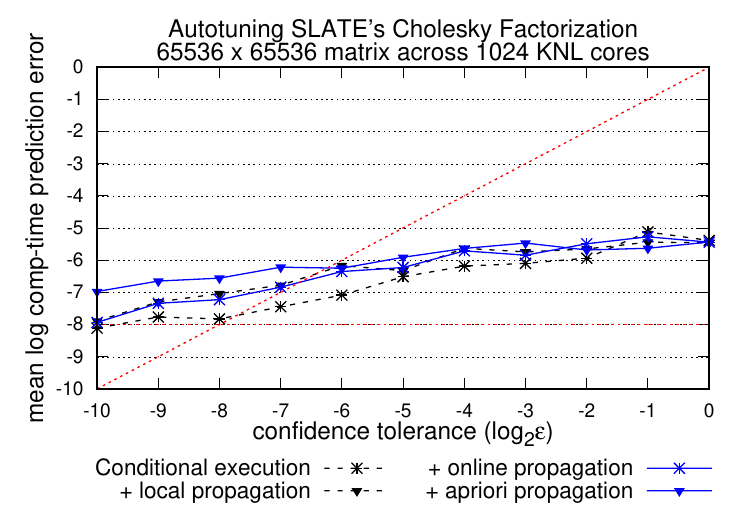}
        \caption{}
        \label{fig:slate_cholesky_comp_err}
      \end{subfigure}
      \begin{subfigure}{0.24\textwidth}
        \centering
        \includegraphics[scale=.6]{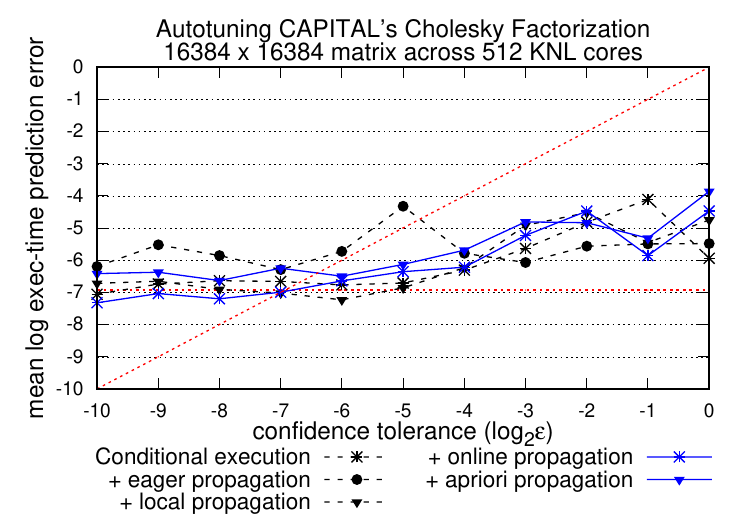}
        \caption{}
        \label{fig:capital_cholesky_err}
      \end{subfigure}
      \begin{subfigure}{0.24\textwidth}
        \centering
        \includegraphics[scale=.6]{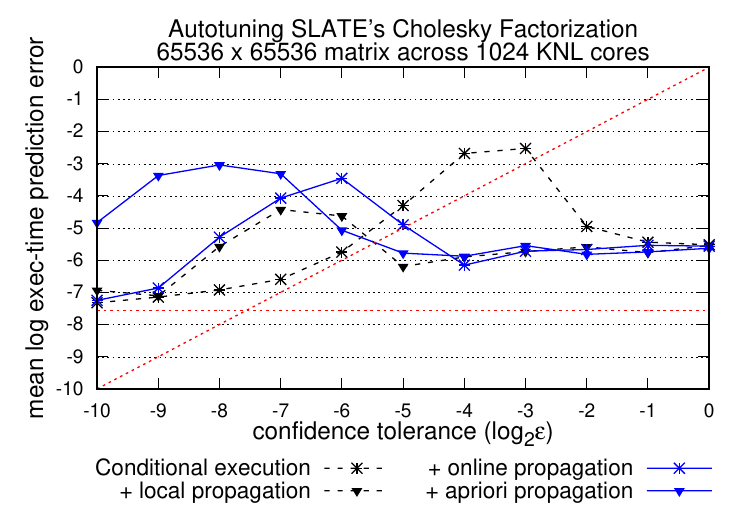}
        \caption{}
        \label{fig:slate_cholesky_err}
      \end{subfigure}
      \begin{subfigure}{0.24\textwidth}
        \centering
        \includegraphics[scale=.6]{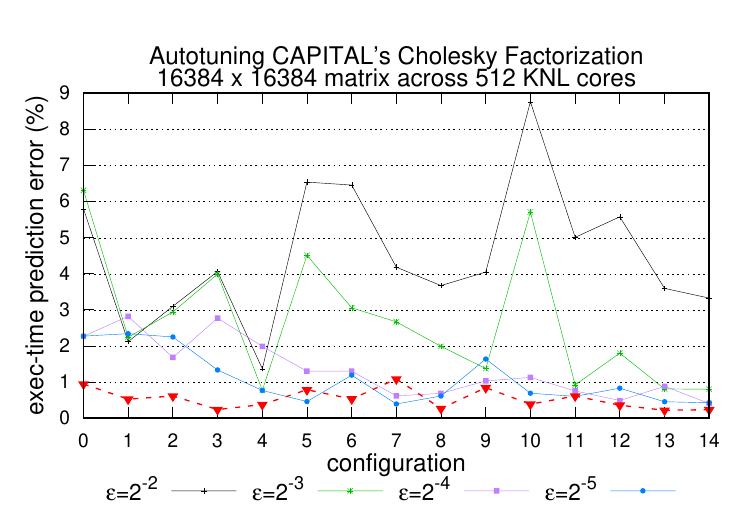}
        \caption{}
        \label{fig:capital_cholesky_config_err}
      \end{subfigure}
      \begin{subfigure}{0.24\textwidth}
        \centering
        \includegraphics[scale=.6]{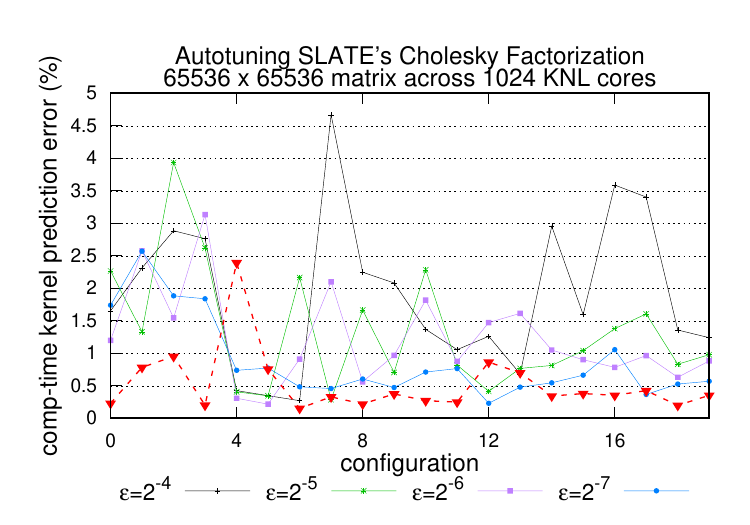}
        \caption{}
        \label{fig:slate_cholesky_comp_config_err}
      \end{subfigure}
      \caption[]{Autotuning execution time and prediction error of approximate autotuning methods for Cholesky factorization. Execution time and prediction error achieved with full kernel execution correspond to red plot lines. Ideal prediction error scaling is represented by diagonal red plot lines. Figures \ref{fig:capital_cholesky_config_err} and \ref{fig:slate_cholesky_comp_config_err} evaluate \textit{online freq propagation}.}
      \label{fig:cholesky_tune_1}
    \end{figure*}

  \section{Approximate autotuning evaluation}

    We use the Stampede2 supercomputer at Texas Advanced Computing Center (TACC)\cite{Stanzione:2017:SEX:3093338.3093385}.
    Stampede2 consists of 4200 Intel Knights Landing (KNL) compute nodes (each capable of a performance rate over 3 Teraflops/s) connected by an Intel Omni-Path (OPA) network with a fat-tree topology (achieving an injection bandwidth of 12.5 GB/sec).
    Each KNL compute node provides 68 cores with 4 hardware threads per core.
    We use 64 MPI processes per node, with 1 thread per process, for all experiments. 
    All implementations of Cholesky and QR factorization use the Intel/18.0.2 environment with MPI version impi/18.0.2.
    Each uses optimization flag -03 and 512-bit vectorization via flag -xMIC-AVX512.
    We utilize parallel MKL BLAS/LAPACK routines via flag -mkl=parallel. 
    All input matrices are generated randomly using double precision.\par

  \subsection{Measurement environment}
  We evaluate the practicality of Critter using the following metrics: relative prediction error for each configuration, mean relative prediction error across all configurations, and autotuning speedup across the configuration space. 
  We additionally evaluate Critter's choice of the optimal configuration. As our framework can be applied to accelerate any configuration-space search strategy, we use exhaustive search to evaluate the efficiency of Critter and the prediction accuracy attained by each method described in Section \ref{critter_sec}.\par

    As Stampede2 does not allocate a contiguous set of nodes, variability in execution time is observed to be high. 
    We measure error of the execution time estimated for a configuration whose kernels are executed selectively by comparing it to a full execution of the configuration.
    This full execution is performed directly prior to the approximated one to minimize variability in the environment.
    To quantify noise level, we measure the variance of five full execution samples. 
    We run each experiment on two distinct node allocations and execute each configuration five times.
    For all methods except \textit{eager propagation}, we execute each kernel at least once per tuning iteration. 
    We reset the performance statistics of all kernels before tuning a new configuration of Slate's and CANDMC's algorithms. 
    Each dense input matrix is reset prior to executing a LAPACK routine to account for error caused by selective kernel execution. 
    All experiments use a 95\% confidence level to construct kernel execution time confidence intervals based on the (scaled) sample variance.\par

    \begin{figure*}[htbp]
      \centering
      \begin{subfigure}{0.24\textwidth}
        \centering
        \includegraphics[scale=.6]{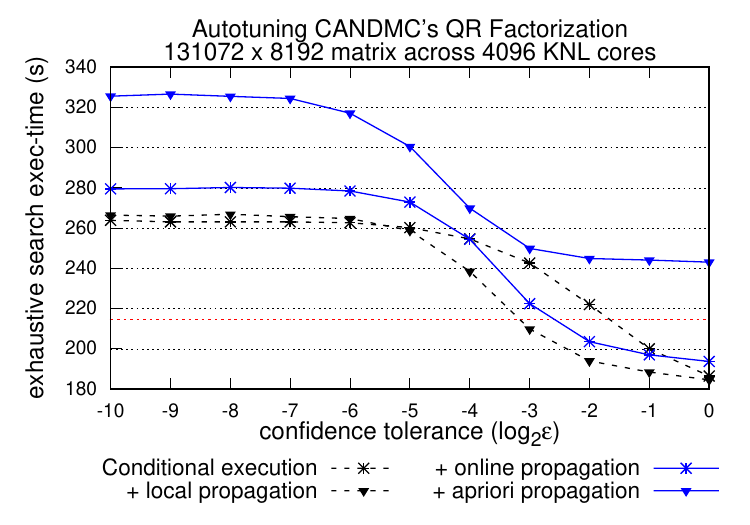}
        \caption{}
        \label{fig:candmc_qr_tt}
      \end{subfigure}
      \begin{subfigure}{0.24\textwidth}
        \centering
        \includegraphics[scale=.6]{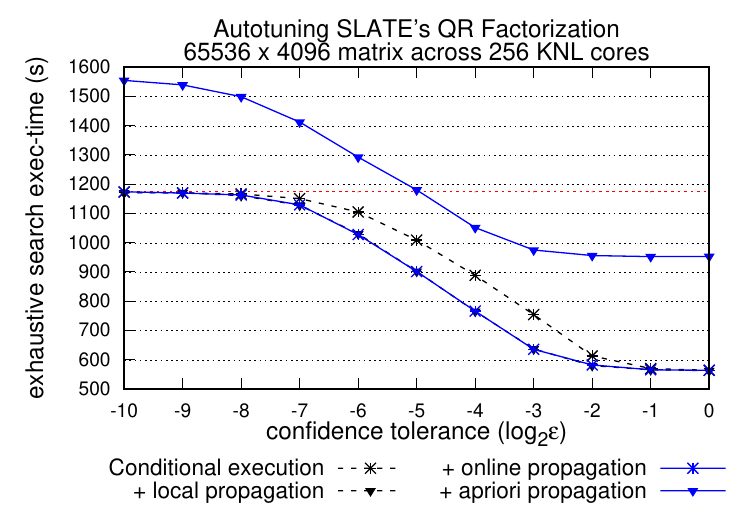}
        \caption{}
        \label{fig:slate_qr_tt}
      \end{subfigure}
      \begin{subfigure}{0.24\textwidth}
        \centering
        \includegraphics[scale=.6]{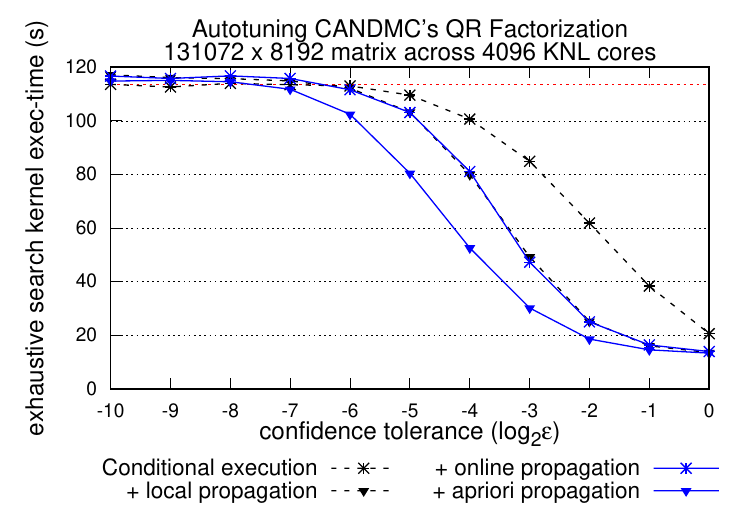}
        \caption{}
        \label{fig:candmc_qr_ktime}
      \end{subfigure}
      \begin{subfigure}{0.24\textwidth}
        \centering
        \includegraphics[scale=.6]{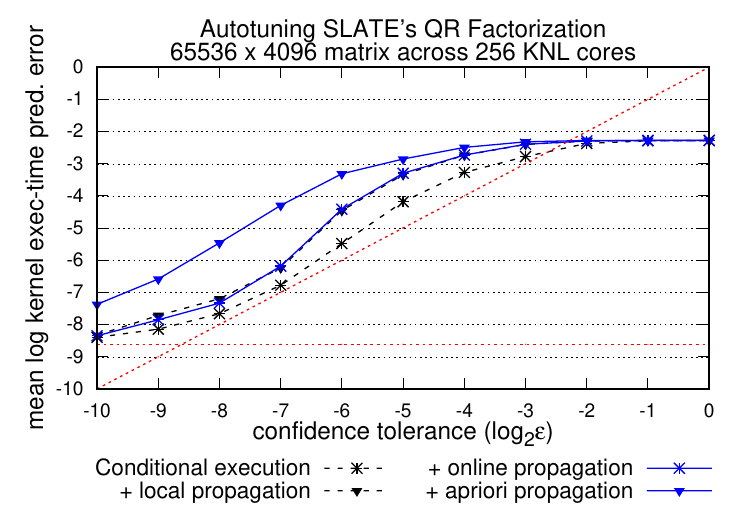}
        \caption{}
        \label{fig:slate_qr_kerr}
      \end{subfigure}
      \begin{subfigure}{0.24\textwidth}
        \centering
        \includegraphics[scale=.6]{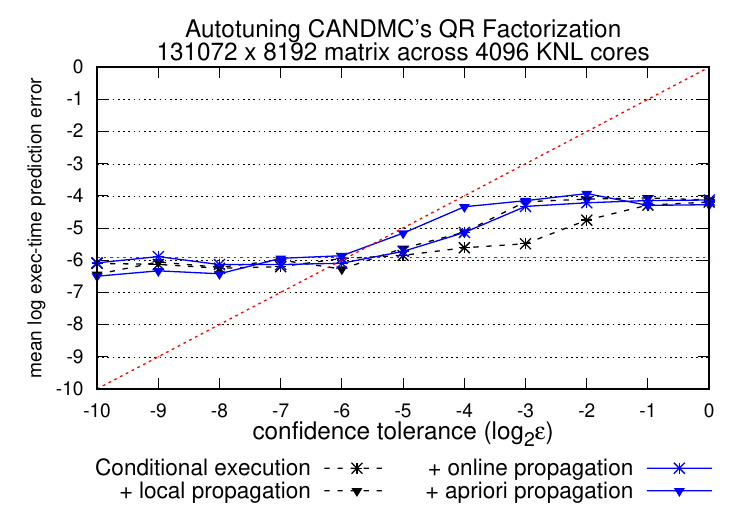}
        \caption{}
        \label{fig:candmc_qr_err}
      \end{subfigure}
      \begin{subfigure}{0.24\textwidth}
        \centering
        \includegraphics[scale=.6]{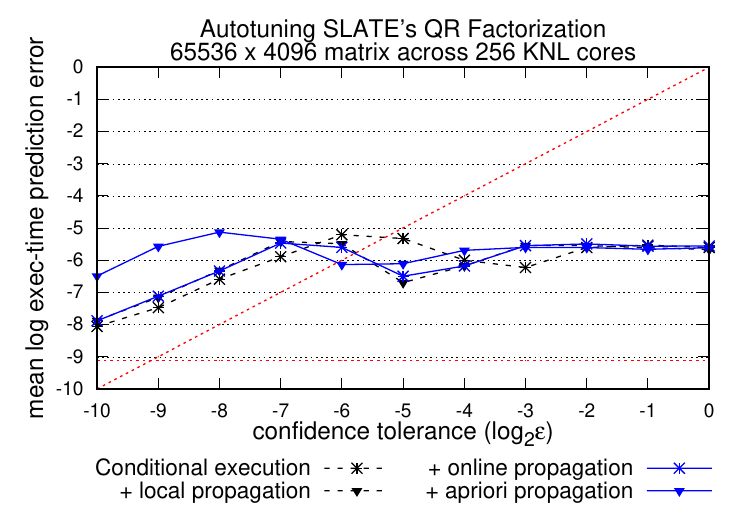}
        \caption{}
        \label{fig:slate_qr_err}
      \end{subfigure}
      \begin{subfigure}{0.24\textwidth}
        \centering
        \includegraphics[scale=.6]{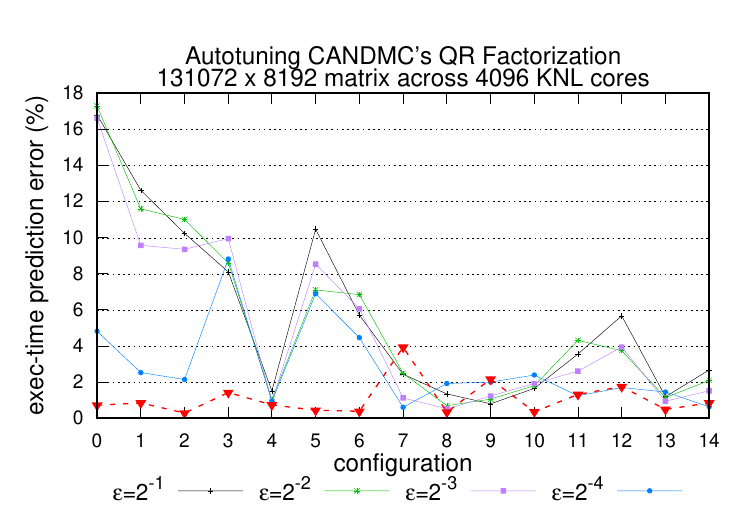}
        \caption{}
        \label{fig:candmc_qr_config_err}
      \end{subfigure}
      \begin{subfigure}{0.24\textwidth}
        \centering
        \includegraphics[scale=.6]{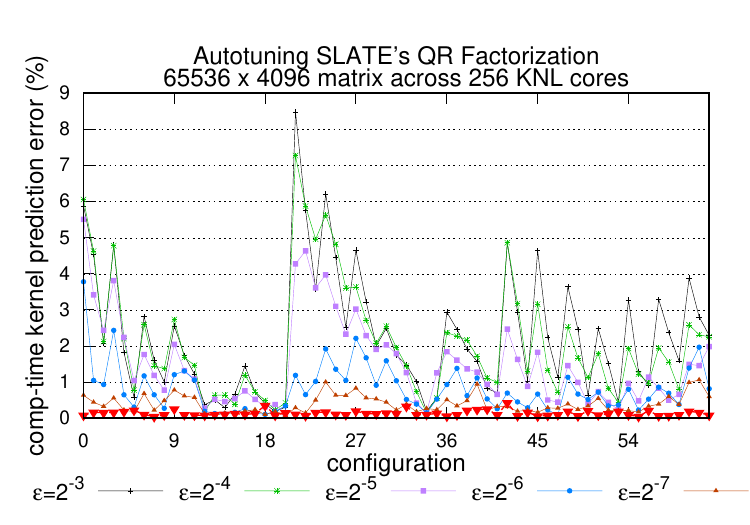}
        \caption{}
        \label{fig:slate_qr_comp_config_err}
      \end{subfigure}
      \caption[]{Autotuning execution time and prediction error of approximate autotuning methods for QR factorization. Execution time and prediction error achieved with full kernel execution correspond to red plot lines. Ideal prediction error scaling is represented by diagonal red plot lines. Figures \ref{fig:candmc_qr_config_err} and \ref{fig:slate_qr_comp_config_err} evaluate \textit{online freq propagation}.}
      \label{fig:qr_tune_1}
    \end{figure*}

  \subsection{Autotuning speedup}

    Figure~\ref{fig:capital_cholesky_tt} shows that selective execution for Capital's Cholesky can lower autotuning time by up to 7.1x, given a sufficiently large prediction error threshold. 
    The frequency with which Capital's kernels are executed depends exponentially on the recursion depth at which they are invoked. 
    Consequently, there are a few BLAS kernels on large matrices and many small BLAS and LAPACK kernels on small matrices. 
    For looser confidence tolerances, speedup from selective execution is limited by the expensive BLAS kernels, as well as communication collectives sending significant amounts of data. 

    For Capital's Cholesky, in Figure~\ref{fig:capital_cholesky_tt}, speed-ups of up to about 1.2x are attained due to propagation of kernel execution counts (\textit{local propagation} and \textit{online propagation} relative to \textit{conditional execution}). 
    \textit{A priori propagation}'s autotuning time incurs an extra full execution of each configuration to determine the confidence interval size necessary to stop kernel execution. This overhead prevents any speedup relative to \textit{conditional execution}. 

    The \textit{eager propagation} method achieves speedups of 2.4 - 7.1x relative to \textit{conditional execution} in Figure \ref{fig:capital_cholesky_tt} without propagating kernel execution counts. 
    These speedups at the looser confidence tolerances are explained by the fact that, unlike other strategies, \textit{eager propagation}  does not require executing each kernel at least once per tuning iteration. 
    This method's aggressive selective execution policy observes a relatively small rate of growth of speed-up as confidence tolerances decrease, which consequently yields speedups at the smaller confidence tolerances. 
    These results indicate that for bulk-synchronous algorithms without communication-computation overlap, reusing kernel performance models across multiple configurations can yield significant speedups relative to both \textit{conditional execution} and full kernel execution.
    \par

    Slate's implementation of Cholesky employs fixed tile sizes, executing BLAS kernels for the same input size many times for a particular configuration, but for other sizes in configurations with different parameters. 
    Figure \ref{fig:slate_cholesky_comp_tt}  
    shows that the longest time any processor spends in executing kernels (excluding profiling overheads) speedup for Slate Cholesky is reduced by up to 75x via selective execution. 
    Speedups in total execution time for each kernel execution count propagation method relative to \textit{conditional execution} in Figure \ref{fig:slate_cholesky_tt} are less significant (up to 1.8x) than those observed in Figure \ref{fig:slate_cholesky_comp_tt} due to computational overhead present when the granularity at which tasks execute is fine (i.e., small tile dimensions). 

    Both the Slate and CANDMC QR algorithms execute similar kernels with many distinct input sizes (i.e., matrix dimensions and message sizes) compared to the Cholesky factorization algorithms.
    Slate's QR factorization implementation exhibits autotuning speedups similar to those of Slate's implementation of Cholesky.
    This is because computational overhead (e.g., BLAS 2 routines that are not executed selectively), prevalent in Slate's small-tile configurations benchmarked in Figure \ref{fig:slate_qr_tt}, diminishes overall speed-up obtained from selective execution. 
    We remark that Critter's profiling overhead is minimal, despite the many messages communicated 
    along the critical path by these QR algorithms (see Figures \ref{fig:slate_cholesky_comm_synch_vary} and \ref{fig:slate_qr_comm_synch_vary}), which reflects its efficiency when profiling nonblocking communications.\par

    Figure \ref{fig:candmc_qr_tt} demonstrates that overall speed-up from selective execution for CANDMC QR is limited to 1.2x.
    CANDMC QR executes kernels for a gradually shrinking trailing matrix size, resulting in a large number of kernels that are modeled independently.
    However, when restricting attention to kernel execution times on the most loaded processor (Figure~\ref{fig:candmc_qr_ktime}), \textit{conditional execution} attains up to 6.6x speedup in kernel execution time relative to full kernel execution. 
    Further, in Figure~\ref{fig:candmc_qr_ktime}, we observe that propagation of kernel execution counts reduces the maximum kernel execution time on any processor by 3.3x relative to \textit{conditional execution}.
    The differences in full execution timings for overall CANDMC QR execution time in Figure~\ref{fig:candmc_qr_tt} and that of the selectively executed kernels in Figure~\ref{fig:candmc_qr_ktime} suggest that speed-up is limited by bottlenecks in CANDMC QR that are not encapsulated in selectively executed kernels.

  \subsection{Prediction error}


    We consider the effect of decreasing the confidence tolerance $\epsilon$ 
    on the observed overall execution time prediction error in Figure~\ref{fig:candmc_qr_config_err} (CANDMC QR) and Figure~\ref{fig:slate_qr_comp_config_err} (Slate QR). 
    We observe that the \textit{conditional execution} method, which does not leverage execution counts to adjust the confidence interval size, achieves superior prediction accuracy relative to the propagation methods for a fixed confidence tolerance $\epsilon$.
  Strategies that propagate critical-path execution counts to perform fewer kernel executions incur more error, but this error still decreases systematically.
  For CANDMC QR (Figure~\ref{fig:candmc_qr_err}), we observe that the error achieves the desired tolerance, while for Slate QR (Figure \ref{fig:slate_qr_err}), the error in the predicted performance exceeds the desired tolerance. 

    We highlight two specific results. The mean prediction error in Slate \texttt{geqrf}'s critical-path kernel execution time in Figure \ref{fig:slate_qr_kerr} is about $1\%$ between $\epsilon=1$ and $\epsilon=2^{-3}$. As the number of kernel invocations increases between $\epsilon=2^{-4}$ and $\epsilon=2^{-8}$, the mean prediction error reduces to less than $.3\%$ for all methods. 

    Figure \ref{fig:slate_qr_comp_config_err} shows the reduction in execution time prediction error achieved by lowering tolerance, across different configurations of Slate QR tuning parameters. 
    The mean prediction error in CANDMC's execution time in Figure \ref{fig:candmc_qr_err} is about $6\%$ between $\epsilon=1$ and $\epsilon=2^{-2}$. As the number of kernel executions increase between $\epsilon=2^{-3}$ and $\epsilon=2^{-6}$, the prediction accuracy increases gradually for all propagation methods. 
    The improvement in performance prediction accuracy is evident for each of the configurations when comparing large confidence tolerances to tolerance $\epsilon=2^{-4}$ in Figure \ref{fig:candmc_qr_config_err}.\par

    Evaluation of 
    Figures \ref{fig:capital_cholesky_err} and \ref{fig:slate_cholesky_comp_err} evaluate the mean prediction error in execution time and critical-path computation time of both Cholesky factorization algorithms in Capital and Slate, respectively.
    Figure \ref{fig:slate_cholesky_comp_err} shows that computational kernel time can be predicted with high accuracy and systematically improved by collecting more kernel execution time samples.
    In particular, Slate's mean prediction error in critical-path computation kernel time decreases from $4\%$ at $\epsilon=1$ to $0.3\%$ at the smallest confidence tolerance.
    Figure~\ref{fig:slate_cholesky_comp_config_err} shows the achieved accuracy for each configuration of tuning parameters, demonstrating that accuracy improvements are achieved for all configurations.

  The nonblocking communication kernels invoked by Slate's algorithms are unpredictable due to variation in the communication-computation overlap present during the execution of each kernel. 
  Systematic reduction in prediction error of execution time is therefore not observed for either algorithm in Figures \ref{fig:slate_cholesky_err} and \ref{fig:slate_qr_err} until $\epsilon$ is sufficiently small.
  However, systematic reduction in prediction error of kernel execution time is observed for Slate Cholesky (Figures~\ref{fig:slate_cholesky_comp_err} and~\ref{fig:slate_cholesky_comp_config_err}) and QR (Figures~\ref{fig:slate_qr_kerr} and~\ref{fig:slate_qr_comp_config_err}).
\par

  Critter correctly selects the optimal QR factorization algorithm configuration for all confidence tolerances $\epsilon$, and selects a configuration for each Cholesky algorithm that achieves at least $99\%$ of the optimal configuration's performance for all $\epsilon$. We remark that these results ameliorate prediction error 
  observed with smaller confidence tolerances, and are not adversely affected by the presence of communication-computation overlap in SLATE's algorithms.

  These results demonstrate that Critter, and the approximate autotuning framework it implements, can accelerate distributed-memory autotuning in a noisy environment while accurately predicting performance of bulk-synchronous algorithms. They also demonstrate that for task-based algorithms, collection of execution time samples can generally systematically reduce prediction error.
  We also observe that Critter can predict certain other metrics, such as the largest time spent in computational kernels along any execution path, with an even better accuracy. 

\section{Related works}
  A number of studies have used autotuning and critical-path analysis to optimize dense linear algebra algorithms. We believe this study is the first to leverage critical-path analysis to predict performance and accelerate autotuning.

  \subsection{Critical-path analysis}

    Critical-path analysis has been used extensively to analyze algorithms in theory, and libraries exist that measure along the critical path~\cite{10.1145/238020.238024,tallent2017representative,bohme2012scalable}.
    Alternative path metrics, including communication and synchronization costs, can characterize execution paths and find use in predicting scaling bottlenecks~\cite{tallent2017representative,chen2015critical}. Tracking a distribution of $k$ paths further enhances the utility of path analysis by quantifying load imbalance, a technique shown to be efficient for moderately large $k$~~\cite{tallent2017representative}. Quantification of critical-path communication costs by profiling MPI communication has been done manually in previous studies on communication-avoiding algorithms~\cite{europar2011,DGKSY_IPDPS_2013}. Critter automates the process used in these papers.

  \subsection{Autotuning mechanisms}

  Autotuning is a widely used technique for the optimization of performance-sensitive kernels of many applications~\cite{balaprakash2018autotuning,10.1145/2628071.2628092}.
  Early dense linear algebra efforts to employ autotuning include the PHiPAC~\cite{bilmes1997optimizing} and the ATLAS library~\cite{whaley1998automatically,whaley2001automated}.
  Acceleration of autotuning for dense linear algebra by pruning the subspace of slow or invalid configuration has been the subject of previous study~\cite{6122021,10.1145/3293320.3293334}.
  Such techniques can be combined with and likely augmented using the profiling and modeling statistics collected online by Critter.
  We note that in many of the above studies, as well as in more complex dense linear algebra algorithms with multiple levels of blocking~\cite{SD_EUROPAR_2011,ballard2014reconstructing}, a much larger tuning space is often used than in the studies performed in this paper.
  In such settings Critter's mechanisms for selective execution have further potential to improve performance.

  Beyond dense linear algebra, autotuning has also been applied extensively in various other applications, such as sparse matrix computations~\cite{byun2012autotuning,guo2010auto,choi2010model}, FFT kernels~\cite{frigo1998fftw,nukada2009auto}, and digital signal processing~\cite{franchetti2018spiral}.
  These include highly successful efforts such as Oski, FFTW, and Spiral, in the three respective domains.
  Online autotuning focused on modeling symmetries and redundancies among configuration parameters has been studied to improve search convergence speed~\cite{pfaffe2019efficient}.
  Previous autotuning work has also developed techniques for automatic learning of performance models for kernels to do offline runtime prediction~\cite{luo2015fast}. 
  We additionally highlight parallel work on improving the cost and accuracy of empirical performance modeling that utilizes compiler-based analysis to selectively execute loop iterations.\cite{copik2020extracting}.
  Our work is set apart from this previous autotuning research by the use of online profiling, selective execution of subroutines, and distributed-memory critical-path analysis.

\section{Conclusion}
  This work presents the first empirical evidence that selective execution of computation and communication kernels can accelerate automatic performance tuning of distributed-memory dense linear algebra libraries.
  It also demonstrates the efficacy of our statistical framework in achieving tunable performance prediction for bulk-synchronous algorithms, and separately in achieving tunable prediction accuracy of critical-path computation time for task-based schedules.
  However, our methods are unable to account for overlap in computation and communication, and cannot quantify the effects of shared resource consumption on kernel performance in the presence of asynchronous kernel execution decisions.\par
  Our techniques should be extensible to other applications and autotuning methods. Automatic performance tuning of dense tensor computations and matrix computations with a fixed sparsity structure in particular can benefit. However, significant practical challenges remain for their application to computations that are different from dense linear algebra. 
  In particular, Critter's selective execution methods are incompatible with matrix computations with varying sparsity structure or other data-dependent computations.\par
  Extensions to our techniques are necessary to further accelerate automatic performance tuning within the proposed framework. Extrapolation of individual kernel performance models to characterize kernel performance across varying input sizes can benefit a wide class of algorithms, including CANDMC's pipelined QR factorization algorithm. Such line-fitting approaches can permit kernel execution to be more selective. 
  Further improvements to our model may also pursue taking into account perturbations to each kernel's execution environment that originate from variability in cache effects or shared resource consumption induced by selective kernel execution.\par

\section*{Acknowledgments}
The first author would like to acknowledge the Department of Energy (DOE) and Krell Institute for support via the DOE Computational Science Graduate Fellowship (grant No.\ DE-SC0019323).
This work used the Extreme Science and Engineering Discovery Environment (XSEDE), which is supported by National Science Foundation grant number ACI-1548562. Via XSEDE, the authors made use of the TACC Stampede2 supercomputer (allocation TG-CCR180006). 
This research has also been supported by funding from the National Science Foundation (grant No.\ 2028861).\par

\bibliographystyle{IEEEtran}
\bibliography{paper}

\end{document}